\documentclass[12pt]{article}%
\usepackage{amssymb}
\usepackage{amsfonts}
\usepackage{sw20elba}%
\usepackage{amsmath}%
\usepackage{graphicx}
\providecommand{\U}[1]{\protect\rule{.1in}{.1in}}

\begin{document}

\title{Superconductivity at 17 K in Yttrium Metal under Nearly Hydrostatic Pressures
to 89 GPa}
\author{J.J. Hamlin$^{a}$, V.G. Tissen$^{b},$ and J.S. Schilling$^{a}\vspace{0.4cm}$\\$^{a}$\textit{Department of Physics, Washington University}\\\textit{\ CB 1105, One Brookings Dr, St. Louis, MO 63130}\\\textit{\ }$^{b}$\textit{Institute of Solid State Physics}\\\textit{\ Chernogolovka 142432, Moscow District, Russia}}
\date{January 10, 2006}
\maketitle

\begin{abstract}
In an experiment in a diamond anvil cell utilizing helium pressure medium,
yttrium metal displays a superconducting transition temperature which
increases monotonically from $T_{c}\simeq$ 3.5 K at 30 GPa to 17 K at 89.3
GPa, one of the highest transition temperatures for any elemental
superconductor. The pressure dependence of $T_{c}$ differs substantially from
that observed in previous studies under quasihydrostatic pressure to 30 GPa.
Remarkably, the dependence of $T_{c}$ on relative volume $V/V_{o}$ is linear
over the entire pressure range above 33 GPa, implying that higher values of
$T_{c}$ are likely at higher pressures. For the trivalent metals Sc, Y, La, Lu
there appears to be some correlation between $T_{c}$ and the ratio
$r_{a}/r_{c}$ of the Wigner-Seitz radius to the ion core radius.

\end{abstract}

\newpage

Before the advent of high-temperature superconductivity in 1986, the highest
known values of the superconducting transition temperature were exhibited by
the binary A-15 compounds V$_{3}$Si, Nb$_{3}$Sn and Nb$_{3}$Ge with $T_{c}$'s
in the range 17 - 23 K \cite{a15ref}. With the discovery in 2001 of
superconductivity in MgB$_{2},$ the highest value of $T_{c}$ for a binary
compound was extended to 40 K \cite{nagamatsu1}. For elemental
superconductors, on the other hand, the maximum value of $T_{c}$ at ambient
pressure is only 9.5 K for Nb. Under high pressure conditions, however, the
number of elemental superconductors not only increases from 29 to 52
\cite{hemleyashcroft,eremets1,schilling1}, but the transition temperatures for
a number of elements (Li, P, S, Ca, V, La) reach values in the range 13 - 20 K
formerly \textquotedblleft reserved\textquotedblright\ for the A-15 compounds
\cite{buzea1}.

In this paper we focus our attention on the four closely related $d$-band
metals Sc, Y, La, and Lu which share the trivalent valence electron
configuration $nd^{1}(n+1)s^{2}$, where $n=3,4$ or 5. Whereas La is
superconducting at ambient pressure, Sc, Y, and Lu only superconduct under
high pressure, all four elements exhibiting a positive pressure derivative
$dT_{c}/dP>0$ \cite{probst1,wittig3,wittig2,tissen1}. Some light on the origin
of these and other interesting results was shed by the observation of
Johansson and Rosengren \cite{johansson1} in 1975 that the ratio of the
Wigner-Seitz radius to ionic radius, $r_{a}/r_{c},$ appears to play an
important role in determining the pressure dependence of the superconducting
properties of Y, La, Lu and La-Y, La-Lu alloys as well as the equilibrium
crystal structure sequence ($hcp\rightarrow$ Sm-type $\rightarrow
dhcp\rightarrow fcc)$ across the rare-earth series. Duthie and Pettifor
\cite{duthie1} subsequently demonstrated that these and other important
correlations are a consequence of the fact that the ratio $r_{a}/r_{c}$ is
inversely related to the $d$-band occupancy $n_{d}$, a quantity which in
general increases under pressure due to $s\rightarrow d$ transfer. Later
studies show that yttrium metal follows the structure sequence
($hcp\rightarrow$ Sm-type $\rightarrow dhcp\rightarrow$ trigonal) as the
applied pressure is increased to 50 GPa at ambient temperature
\cite{vohra1,grosshans1}. Melsen \textit{et al.} \cite{melsen1} have predicted
that at pressures above 280 GPa yttrium should transform into the $bcc$
structure. It would be of great interest to extend the above
superconductivity/structural experiments to much higher pressures to allow a
critical assessment of possible correlations with the ratio $r_{a}/r_{c}$ over
a wide range of parameters.

Yttrium metal does not superconduct above 6 mK at ambient pressure
\cite{probst1}. However, in 1970 Wittig \cite{wittig1} discovered
superconductivity in Y at $T_{c}\simeq$ 1.3 K under 11 GPa quasihydrostatic
pressure (solid steatite pressure medium); $T_{c}$ increases monotonically
with pressure at the rate $dT_{c}/dP\simeq+0.35$ K GPa$^{-1},$ finally
reaching 9 K at 30 GPa. In the present paper we extend these earlier studies
to much higher pressures; we also provide for a nearly hydrostatic pressure
environment by using dense helium as pressure medium. We find that $T_{c}$
indeed increases monotonically with pressure, ultimately reaching 17 K at 89.3
GPa. This is one of the highest values of $T_{c}$ ever observed for an
elemental superconductor; values above 20 K appear likely at higher pressures.
Comparing the pressure dependences for Y, Lu, La, and Sc, the simple inverse
relation between $T_{c}$ and the ratio $r_{a}/r_{c}$ proposed by Johansson and
Rosengren \cite{johansson1} is found to extend to much higher pressures.

High pressures were generated using a diamond anvil cell (DAC) made of CuBe
alloy, nonmagnetic CuBe being used in the critical region near the sample
\cite{albany1}. Two opposing 1/6-carat type IIa diamond anvils with 0.3 mm dia
culet and 3 mm table were used. A miniature Y sample was cut from an ingot
(Aldrich Chemical 99.9\%) to approximate dimensions $60\times60\times20$ $\mu
$m$^{3}$ and placed in a 150 $\mu$m dia hole electro-spark-drilled through the
center of a 3 mm dia $\times$ 250 $\mu$m thick gold-plated rhenium or
NiCrAl-alloy gasket preindented to 50 $\mu$m (see Fig 1 in Ref \cite{deemyad2}%
). The rhenium gasket used in experimental runs $A$ and $B$ becomes
superconducting near 3.5 K under pressure \cite{deemyad2,chu5}, thus allowing
the detection of the superconducting signal from the Y sample only for
$T_{c}\geq$ 4 K. For this reason a nonsuperconducting gasket made of
NiCrAl-alloy was used in run $C$.

Tiny ruby spheres \cite{note1} allow the determination \cite{mao1} of the
pressure \textit{in situ} with resolution $\pm$ 0.2 GPa at 20 K. For the
results shown here, the standard\ ruby calibration in Ref \cite{mao1} was
used; however, we point out that Holzapfel \cite{holzapfel1} has very recently
published a revised ruby pressure calibration to 300 GPa which deviates
significantly (%
$>$
5\%) from the previous calibration in the pressure range above 60 GPa.
According to this revised ruby scale our highest pressure should be corrected
upwards from 89.3 GPa to 96 GPa.

At the beginning of the experiment, the gasket hole is filled with superfluid
liquid helium at temperatures below 2 K before sealing it shut by pressing the
opposing diamond anvils further into the preindented gasket. Pressure is
changed in the temperature range 150 - 180 K. To reduce the chance of helium
penetration into the diamond anvils, the DAC was kept at temperatures below
180 K during the entire duration ($\sim$ 10 days) of each of the three
experimental runs.

The superconducting transition is determined inductively using two balanced
primary/secondary coil systems connected to a Stanford Research SR830 digital
lock-in amplifier. The $ac$ susceptibility studies were carried out using a 3
G (r.m.s.) magnetic field at 1000 Hz. As seen in Fig 1 for the data in run
$B$, the real-part of the $ac$ susceptibility signal changes abruptly at the
superconducting transition by 1-2 nV. The relatively low noise level ($\sim$
0.2 nV) is achieved by appropriate signal compensation and impedance matching
as well as through both the use of a long time constant (30 s) on the lock-in
amplifier during very slow (100 mK/min) temperature sweeps and the averaging
of multiple measurements. Further experimental details of the DAC and $ac$
susceptibility techniques are published elsewhere \cite{albany1,deemyad1}.

In Fig 2 the value of $T_{c}$ from the transition midpoint is plotted versus
pressure for all three experimental runs, revealing excellent agreement.
Normally $T_{c}$ is measured for increasing pressure; however, at the end of
run $A$ the pressure was reduced from 48 to 38 GPa (pt 7 to pt 8),
demonstrating the reversibility of the pressure dependence $T_{c}(P),$ at
least in this pressure range. The present results, which were obtained under
nearly hydrostatic pressure conditions, are seen to differ significantly from
those obtained earlier under quasihydrostatic pressure conditions where a
solid (steatite) was used as pressure medium \cite{wittig1,wittig2}. Abrupt
changes in the slope $dT_{c}/dP$ near 12 and 25 GPa in the earlier data, and
at 30-35 GPa in the present data, may be related to the structural transitions
reported near 15 GPa ($hcp\rightarrow$ Sm-type) and 30 GPa (Sm-type
$\rightarrow dhcp$) at ambient temperature \cite{vohra1,grosshans1}. Note that
these phase boundaries may shift upon cooling from ambient to low temperatures.

In Fig 3 we replot the present results from Fig 2 as $T_{c}$ versus relative
volume $V/V_{o}$ using the equation of state for Y determined by Grosshans and
Holzapfel \cite{grosshans1}. Remarkably, over the entire pressure range 33 to
89.3 GPa, $T_{c}$ is seen to be a linear function of the sample volume $V$.
Were this linear dependence to continue, the transition temperature $T_{c}$
would reach values of 20, 25, or 30 K for pressures of approximately 130, 250,
or 540 GPa, respectively.

We now explore in Fig 4 whether the observed increase in $T_{c}$ with pressure
for the four trivalent elements Sc, Y, La, and Lu is correlated with the ratio
$r_{a}/r_{c},$ as originally proposed by Johansson and Rosengren
\cite{johansson1}, where $r_{a}=\sqrt[3]{(3/4\pi)V_{a}(P)},$ $V_{a}(P)$ is the
volume per atom at the given pressure \cite{grosshans1,springer1}, and the
ionic radius $r_{c}$ \cite{springer1} is assumed independent of pressure. In
examining the data in Fig 4 one should keep in mind that the results of
quasihydrostatic pressure studies (solid or dashed lines) are included
together with those of nearly hydrostatic studies (symbols). That the two very
different pressure environments can have a strong influence on the measured
$T_{c}(P)$ dependences is evident from the results on Y in Figs 2 or 4. In
addition, the results on La of Tissen \textit{et al.} \cite{tissen1}\ using
methanol-ethanol pressure medium differ from the earlier, less hydrostatic
studies \cite{wittig2}.

In spite of these caveats, however, two simple systematics are evident in Fig
4. Firstly,\ the three nonsuperconducting metals Y, Lu, and Sc become
superconducting if high pressure is applied, $T_{c}$ generally increasing with
pressure (decreasing ratio $r_{a}/r_{c}$) for all four metals. Secondly, the
value of $T_{c}$ does not increase above 1 K unless the applied pressure is
sufficient to bring the ratio down to values below $r_{a}/r_{c}\simeq$ 2.1.
For La at ambient pressure the ratio $r_{a}/r_{c}$ is clearly less than 2.1;
this is consistent with the fact that La's \textit{dhcp} phase is
superconducting at $T_{c}\simeq$ 5 K and its \textit{fcc} phase at 6 K. It
would be interesting to investigate possible correlations between $T_{c}$ and
the ratio $r_{a}/r_{c}$ for Sc, Y, La, and Lu to nearly hydrostatic (dense He)
pressures well above 100 GPa (1 Mbar) and, in particular, to determine for
what value of $r_{a}/r_{c}$ the transition temperature $T_{c}$ passes through
its maximum value $T_{c}^{\max}$. The value of $T_{c}^{\max}$, and the
pressure (or ratio $r_{a}/r_{c})$ at which it occurs, may depend on the degree
of hydrostaticity of the pressure medium used.

The value of the superconducting transition temperature found here for Y under
89.3 GPa nearly hydrostatic (dense He) pressure, $T_{c}\simeq$ 17 K from the
midpoint of the magnetic susceptibility transition, is among the highest ever
reported for an elemental superconductor. Using the same
susceptibility-midpoint criterium, Ishizuka \textit{et al.} \cite{ishizuka1}
report that $T_{c}\simeq$ 16.5 K for vanadium under 120 GPa nonhydrostatic
pressure (no pressure medium), with a superconducting onset at 17.2 K. The
superconducting onset in the susceptibility of sulfur takes on a value as high
at 17 K at 157 GPa nonhydrostatic pressure \cite{struzhkin10}. Values of
$T_{c}\approx$ 18 K (30 GPa) and 20 K (48 GPa) have been, respectively,
reported for Ca and Li from their resistivity onsets by Shirotani \textit{et
al.} \cite{shirotani1} and Shimizu \textit{et al.} \cite{shimizu1} for
nonhydrostatic pressure; however, it is well known that the temperature of the
resistivity onset may lie significantly higher than the bulk value of $T_{c}$
\cite{lortz1}. Subsequent magnetic susceptibility experiments on Li report
$T_{c}^{\max}\simeq$ 16 K (transition onset) at 33 GPa \cite{struzhkin11} and
$T_{c}^{\max}\simeq$ 14 K (transition midpoint) at 30 GPa \cite{deemyad2}.
Clearly the value of $T_{c}$ depends to some extent on the measurement
technique and $T_{c}$-criterion used. In any case, the highest reported values
of $T_{c}$ for elemental superconductors under extreme pressure lie near 17 K.
It is interesting to note that for the elements Ca \cite{shirotani1}, Y, Lu
\cite{probst1}, Sc \cite{wittig3}, V \cite{ishizuka1}, B \cite{eremets8}, S
\cite{struzhkin10}, and P \cite{shirotani5} the transition temperature $T_{c}$
is still climbing for the highest pressures reached. It is very likely that in
the near future the transition temperature of one of these elemental
superconductors will surpass the $T_{c}=20$ K barrier under extreme pressures.

To our knowledge, no electronic structure calculation of $T_{c}$ has yet been
carried out for Y at reduced lattice parameters. Such a calculation for V to
945 GPa is in quite good agreement with the experimental results to 120 GPa
and predicts that $T_{c}$ should pass through a maximum value of 21 K at 139
GPa \cite{louis1}. In view of the linear dependence of $T_{c}$ on $V/V_{o}$ to
the highest pressure (see Fig 3), it would be of particular interest to carry
out a similar calculation for Y.\emph{\ }

\vspace{0.4cm}\noindent Acknowledgments. \ The authors would like to
gratefully acknowledge research support by the National Science Foundation
through grant DMR-0404505. Thanks are due V. Struzhkin for providing the NiCrAl-alloy.

\newpage

\begin{center}
{\LARGE Figure Captions}

\bigskip

\bigskip
\end{center}

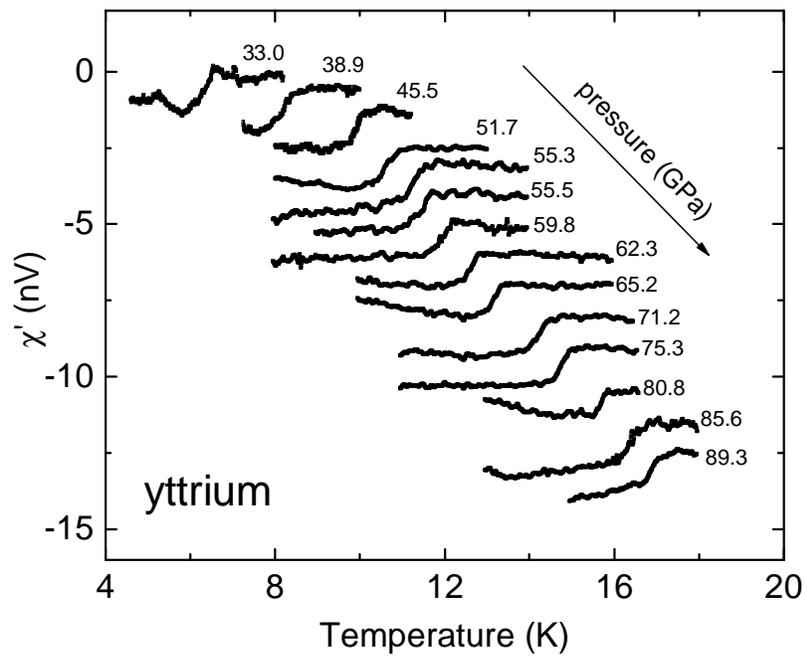
\noindent\textbf{Fig. 1. }Real part of the $ac$ susceptibility signal in
nanovolts versus temperature for yttrium metal at 14 different pressures from
33 to 89.3 GPa (run $B$). The applied $ac$ field is 3 G (r.m.s.) at 1,000 Hz.
Data at different pressures are shifted vertically for clarity. The 1-2 nV
jump in the $ac$ susceptibility marks the superconducting transition at
$T_{c}.$ As the pressure increases, $T_{c}$ is seen to increase monotonically.

\bigskip

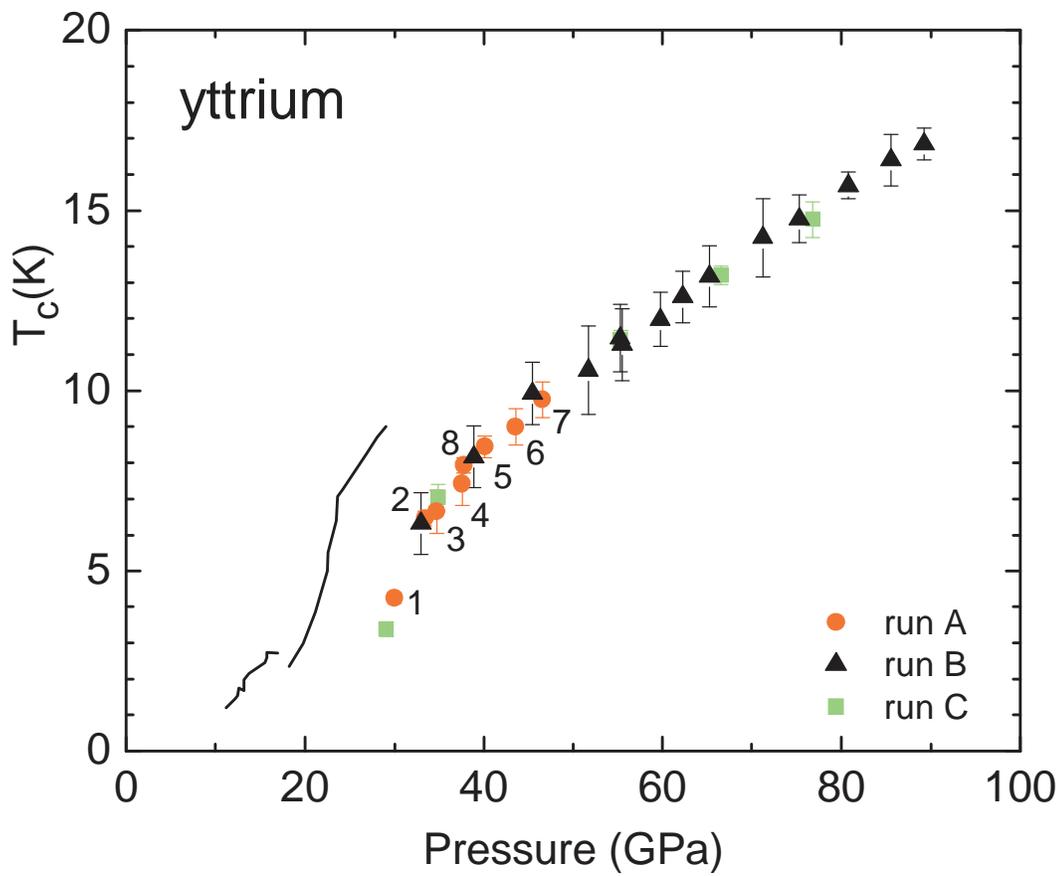
\noindent\textbf{Fig. 2. }Symbols give superconducting transition temperature
of yttrium metal versus nearly hydrostatic (dense helium) pressure to 89.3 GPa
in present experiments. Error bar gives transition width. Numbers give order
of measurements in run $A$; in runs $B$ and $C$ pressure increases
monotonically. Solid lines give $T_{c}(P)$ under quasihydrostatic pressure to
16 GPa from Ref \cite{wittig1} and to 30 GPa from Ref \cite{probst1}.

\bigskip

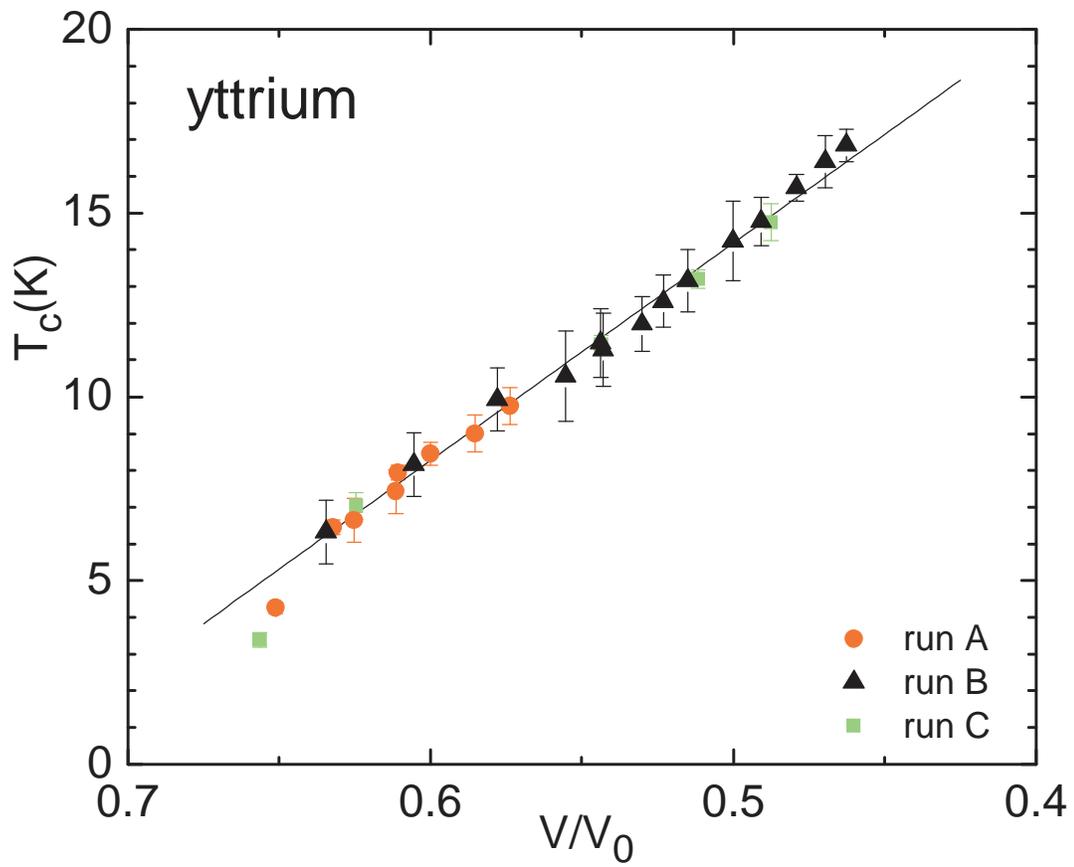
\noindent\textbf{Fig. 3. }Results of present experiments in Fig 2 replotted as
$T_{c}$ versus relative volume $V/V_{o}$ using equation of state from Ref
\cite{grosshans1}. Straight line is drawn to emphasize the linear dependence
for data above 33 GPa: \ $T_{c}($K$)=43.8-59.2(V/V_{o}).$

\bigskip

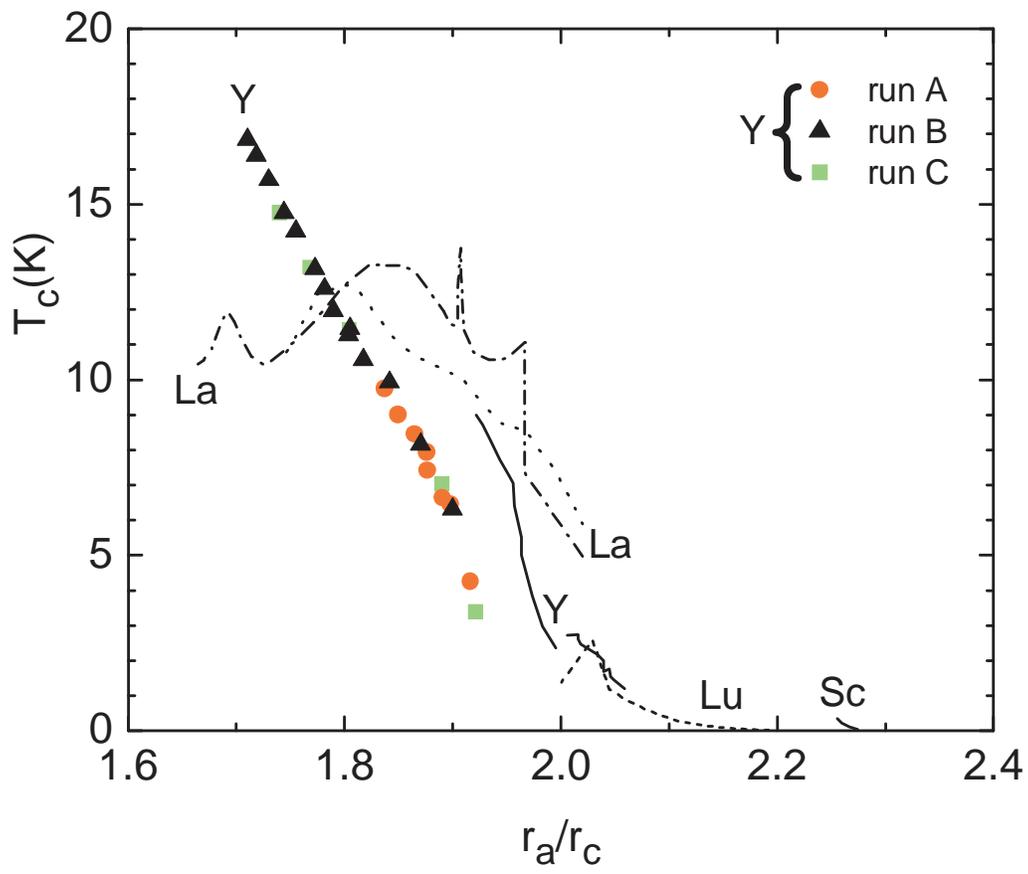
\noindent\textbf{Fig. 4. }Value of $T_{c}$ versus ratio of Wigner-Seitz radius
to trivalent ionic radius, $r_{a}/r_{c},$ for present nearly hydrostatic data
on Y from Fig 2 (symbols) as well as for less hydrostatic data on Y
\cite{probst1} and Sc \cite{wittig3} (solid lines), on La \cite{wittig2} and
Lu \cite{probst1} (dashed lines), and on La from Ref \cite{tissen1}%
\ (dot-dashed lines). At ambient pressure the value of the ratio $r_{a}/r_{c}$
is \cite{springer1}: Y (2.21), Lu (2.23), Sc (2.45), and La (2.08).


\begin{thebibliography}{99}
\bibitem {note1}The ruby spheres, which were annealed at 1500$^{\circ}$C for
one week followed by a cooldown to room temperature over a three week period,
were prepared by J.C. Chervin, Physique des Milieux Condens\'{e}s, Paris, and
contained 3600 ppm Cr$^{3+}$.
\end{thebibliography}
\end{document}